\documentclass[twocolumn,prb]{revtex4}
\usepackage{amsfonts}
\usepackage[T1]{fontenc}
\usepackage{amsmath,amsbsy,amssymb,graphicx}
\usepackage{times}

\begin{document}

\title{Nonlinear non-Hermitian higher-order topological laser}
\author{Motohiko Ezawa}
\affiliation{Department of Applied Physics, University of Tokyo, Hongo 7-3-1, 113-8656,
Japan}

\begin{abstract}
We investigate topological lasers in combination of nonlinear, non-Hermitian
and topological lattice systems based on a quench dynamics starting from one
site. We consider explicitly the topological laser in the
Su-Schrieffer-Heeger (SSH) model with two topological edge states and the
second-order topological laser in the breathing Kagome lattice with three
topological corner states. Once we stimulate any one site, after a delay,
all sites belonging to the topological edge or corner states begin to emit
stable laser light depending on the density of states, although no wave
propagation is observed from the stimulated site. It is intriguing that the
profile of topological edge or corner states is observable by measuring the
intensity of lasing. The phenomenon occurs due to a combinational effect of
linear non-Hermitian loss terms and nonlinear non-Hermitian gain terms in
the presence of the topological edge or corner states.
\end{abstract}

\maketitle

\section{Introduction}

Topological physics is one of the most essential concepts found in recent
fundamental physics\cite{Hasan,Qi}. Recently, it is ubiquitously found in
various systems in photonic\cite%
{KhaniPhoto,Hafe2,Hafezi,WuHu,TopoPhoto,Ozawa16,Ley,KhaniSh,Zhou,Jean,Ota18,Ozawa,Ota19,OzawaR,Hassan,Ota,Li,Yoshimi,Kim,Iwamoto21}%
, acoustic\cite{Prodan,TopoAco,Berto,Xiao,He,Abba,Xue,Ni,Wei,Xue2},
mechanical\cite{Lubensky,Chen,Nash,Paul,Sus} and electric circuit\cite%
{TECNature,ComPhys,Hel,Lu,YLi,EzawaTEC,EzawaLCR,EzawaSkin} systems. Among
them, topological photonics is most extensively studied theoretically and
experimentally. One of the reasons is that it is possible to observe
real-time and real-space dynamics. Another merit is that topological
photonics has opened a new field of topological physics, i.e. non-Hermitian
topology and nonlinear topology and their combination. Non-Hermitian effects
are introduced by a loss and a gain of photons. On the other hand, nonlinear
effects are introduced by the Kerr effect or a stimulated emission effect.

A topological laser is a prominent application of topological physics\cite%
{Harrari,Bandres,Schome,Weimann,Jean,Ota18,Parto,Zhao,Malzard,MalzardOpt,Zhong}%
. It utilizes topological edge states for the coherent laser emission.
Thanks to the topological protection, the topological laser is robust
against the randomness and the defects of the sample, which is favorable for
future laser applications. A topological laser is an ideal play ground to
investigate nonlinear non-Hermitian topological physics. The loss of photons
and gain from stimulated emissions constitute the non-Hermitian terms. The
nonlinear effect is included in the gain term, which represents the
saturation of the gain.

Higher-order topological phases are extension of topological phases\cite%
{Fan,Science,APS,Peng,Lang,Song,Bena,Schin,EzawaKagome}. There emerge
topological corner states in the second-order topological phase. A simplest
example is given by the breathing Kagome lattice, where three topological
corner states appear in triangle geometry. This model is a natural
generalization of the Su-Schrieffer-Heeger (SSH) model to two dimensions.
Since the model requires only positive hoppings, it is realized in various
systems including photonic\cite{Hassan,Li}, acoustic\cite{Xue,Ni} and
electric circuit systems\cite{WuKagome}.

In this paper, we analyze a quench dynamics of a nonlinear-non-Hermitian
topological laser and a higher-order topological laser by stimulating any
one site. We study explicitly the SSH model for a topological laser and the
breathing Kagome model for a higher-order topological laser. In the SSH
model with two topological edge states, once we stimulate any one site,
after a delay, all sites belonging to the two topological edge states begin
to emit stable laser light depending on the density of states (DOS),
although no wave propagation is observed from the stimulated site. Thus, the
profile of a topological edge state together with the DOS is observable by
laser intensity. The strength of the laser light is identical for the two
edges because of reflection symmetry. This is also the case for the
second-order topological laser in the breathing Kagome lattice with three
topological corner states. Here, the system has trigonal symmetry. The
phenomenon occurs due to a combinational effect of linear non-Hermitian loss
terms and nonlinear non-Hermitian gain terms in the presence of the
topological edge or corner states.

\section{Topological laser}

\subsection{Model}

We consider a coupled-ring system made of active resonators\cite{Harrari} .
The dynamics of a laser system is governed by\cite{Harrari} 
\begin{equation}
i\frac{d\psi _{n}}{dt}=\sum_{nm}M_{nm}\psi _{m}-i\gamma \left( 1-\xi \frac{%
P_{n}}{1+\left\vert \psi _{n}\right\vert ^{2}/\eta }\right) \psi _{n},
\label{DSG}
\end{equation}%
where $\psi _{n}$ is the amplitudes of the site $n$, where $n=1,2,3,\cdots
,N $ in the system composed of $N$\ sites; $M_{nm}$ describes a hopping
matrix; $\gamma $ represents the loss in each resonator; $\gamma \xi $
represents the amplitude of the optical gain via stimulated emission; $\eta $
represents the nonlinearity; $P_{n}$ stands for the spatial profile of the
pump. The system turns into the linear model in the limit $\eta \rightarrow
\infty .$ On the other hand, $\gamma $\ controls the non-Hermicity. The
system turns into a Hermitian model for $\gamma =0$. We call the term\
proportional to $\gamma $ the loss term and the term\ proportional to $%
\gamma \xi $\ the nonlinear gain term.

We take%
\begin{equation}
P_{n}=\sum_{\bar{n}}\delta _{n,\bar{n}},
\end{equation}%
where $\bar{n}$ runs over the edge or corner sites. Namely, optical gains
are introduced only at the edge or corner sites.

\begin{figure}[t]
\centerline{\includegraphics[width=0.48\textwidth]{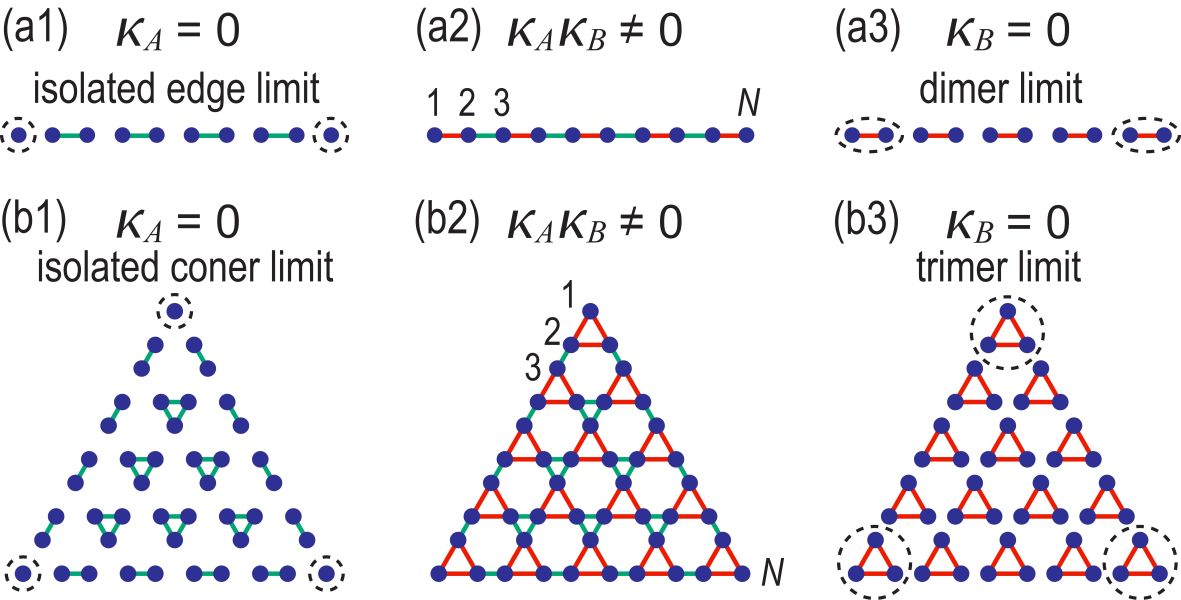}}
\caption{(a1)$\sim $(a3) Illustration of a dimerized lattice with (a1) $%
\protect\kappa _{A}=0$, (a2) $\protect\kappa _{A}\protect\kappa _{B}\neq 0$
and (a3) $\protect\kappa _{B}=0$. (b1)$\sim $(b3) Illustration of a
breathing Kagome lattice with (b1) $\protect\kappa _{A}=0$, (b2) $\protect%
\kappa _{A}\protect\kappa _{B}\neq 0$ and (b3) $\protect\kappa _{B}=0$. A
line (triangle) contains many small segments (triangles). At the edges
(corners) of the chain (triangle), there are two (three) isolated atoms for $%
\protect\kappa _{A}=0$, while there are dimer (trimer) states for $\protect%
\kappa _{B}=0$. They are marked by dotted circles. Lattice sites are
numbered from $n=1$ to $N$ as indicated in (a2) and (b2).}
\label{FigKagomeIllust}
\end{figure}

We are interested in the case where $M_{nm}$ represents a tight-binding
model possessing a topological phase. We explicitly consider the SSH model
illustrated in Fig.\ref{FigKagomeIllust}(a1)$\sim $(a3), where $\bar{n}$\
takes values at the left and right edges, and the breathing Kagome model
illustrated in Fig.\ref{FigKagomeIllust}(b1)$\sim $(b3), where $\bar{n}$\
takes values at the top, bottom-left and bottom-right corners. See Appendix
for the topological charges in these models.

It is possible to solve Eq.(\ref{DSG}) numerically for explicit system
parameters, as we do later. However, to reach a deeper understanding of the
phenomena, an analytical study is indispensable. Since this is impossible
for general system parameters, we make an analytical study for special cases.

\subsection{Edge or corner dynamics}

We first consider the dynamics of an edge or corner site when it is
perfectly isolated as in Fig.\ref{FigKagomeIllust}(a1) or (b1). This is the
case where $M_{nm}=0$\ for the edge or corner sites. For instance, this is
realized by setting $\kappa _{A}=0$\ in Eq.(\ref{HoppiMatrix}) or $\lambda
=-1$ in Eq.(\ref{DimerParam}) for the SSH model.

\begin{figure}[t]
\centerline{\includegraphics[width=0.48\textwidth]{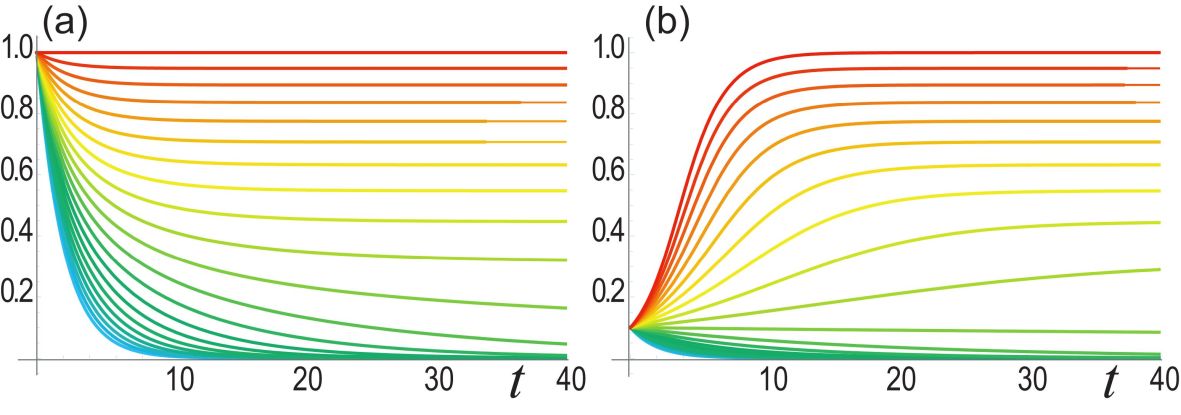}}
\caption{Dynamics of the isolated equation (\protect\ref{NHEq}) under the
initial conditions (a) $\protect\psi _{1}(0)=1$ and (b) $\protect\psi %
_{1}(0)=0.1$. The solutions reach the same stationary solution irrespective
of the initial conditions. Red color indicates the solution with $\protect%
\xi =2$, while cyan color indicates that with $\protect\xi =0$. We have set $%
\protect\eta =1$ and $\protect\gamma =1/2$.}
\label{FigIsolate}
\end{figure}

The dynamics is governed by isolated equations,%
\begin{equation}
\frac{d\psi _{\bar{n}}}{dt}=-\gamma \left( 1-\xi \frac{1}{1+\left\vert \psi
_{\bar{n}}\right\vert ^{2}/\eta }\right) \psi _{\bar{n}}.  \label{NHEq}
\end{equation}%
Solving this equation numerically with the initial conditions $\psi _{\bar{n}%
}=1$ and $\psi _{\bar{n}}=0.1$, we show the results in Fig.\ref{FigIsolate}%
(a) and (b), respectively. It is intriguing that the dynamics of the
topological edge or corner state does not depend on the initial condition of 
$\psi _{\bar{n}}$. The saturated value of $\psi _{\bar{n}}$ as $t\rightarrow
\infty $ is identical for all isolated edge or corner states. This can be
understood analytically as follows.

We solve Eq.(\ref{NHEq}) for nontrivial stationary solutions. The stationary
solution for an edge or corner site $\bar{n}$ is given by%
\begin{equation}
\xi \frac{1}{1+\left\vert \psi _{\bar{n}}\right\vert ^{2}/\eta }=1.
\end{equation}%
Hence, the nontrivial solution reads%
\begin{equation}
\lim_{t\rightarrow \infty }\left\vert \psi _{_{\bar{n}}}\right\vert
^{2}=\eta (\xi -1),  \label{stationary}
\end{equation}%
where it is necessary that $\xi >1$. We have the trivial solution, 
$\lim_{t\rightarrow \infty }\left\vert \psi _{_{\bar{n}}}\right\vert ^{2}=0$,
for $\xi \leq 1$.

As a result, there are only two stable solutions in Eq.(\ref{NHEq}). One is
the ground state mode $\psi _{_{\bar{n}}}=0$, and the other is the
stimulated mode $\left\vert \psi _{_{\bar{n}}}\right\vert ^{2}=\eta \left(
\xi -1\right) $ for $\xi >1$. The initial condition determines which state
is realized.

The nonlinear term is essential to have a stationary solution. Indeed, we
obtain a linear theory in the limit $\eta \rightarrow \infty $\ , where the
amplitude $\left\vert \psi _{n}\right\vert ^{2}$ diverges.

The state remains to be real for a real initial condition, and Eq.(\ref{NHEq}%
) is simplified%
\begin{equation}
\frac{d\psi }{-\psi _{\bar{n}}+\xi \frac{1}{1+\psi _{\bar{n}}^{2}/\eta }\psi
_{\bar{n}}}=\gamma dt,
\end{equation}%
which is solved as%
\begin{equation}
\frac{2\log \psi _{\bar{n}}-\xi \log \left[ \gamma \left( \psi _{\bar{n}%
}^{2}/\eta +1-\xi \right) \right] }{2\left( \xi -1\right) }=\gamma \left(
t+t_{0}\right) .
\end{equation}%
The state evolution $\psi _{\bar{n}}\left( t\right) $ is given by the
inverse of this equation.

\subsection{Bulk dynamics}

We next analyze the dynamics of the bulk site, where there is no nonlinear
non-Hermitian term because $P_{n}=0$. Then, the dynamics is governed by the
linear equation%
\begin{equation}
i\frac{d\psi _{n}}{dt}=\sum_{nm}M_{nm}\psi _{m}-i\gamma \psi _{n}.
\label{BulkEq}
\end{equation}%
With the use of a solution of the linear equation%
\begin{equation}
i\frac{d\psi _{n}^{0}}{dt}=\sum_{nm}M_{nm}\psi _{m}^{0},
\end{equation}%
the solution of Eq.(\ref{BulkEq}) is written in the form%
\begin{equation}
\psi _{n}=e^{-\gamma t}\psi _{n}^{0}.  \label{Decay}
\end{equation}%
It means that the amplitude exponentially decays as a function of time in
the bulk.

\begin{figure}[t]
\centerline{\includegraphics[width=0.48\textwidth]{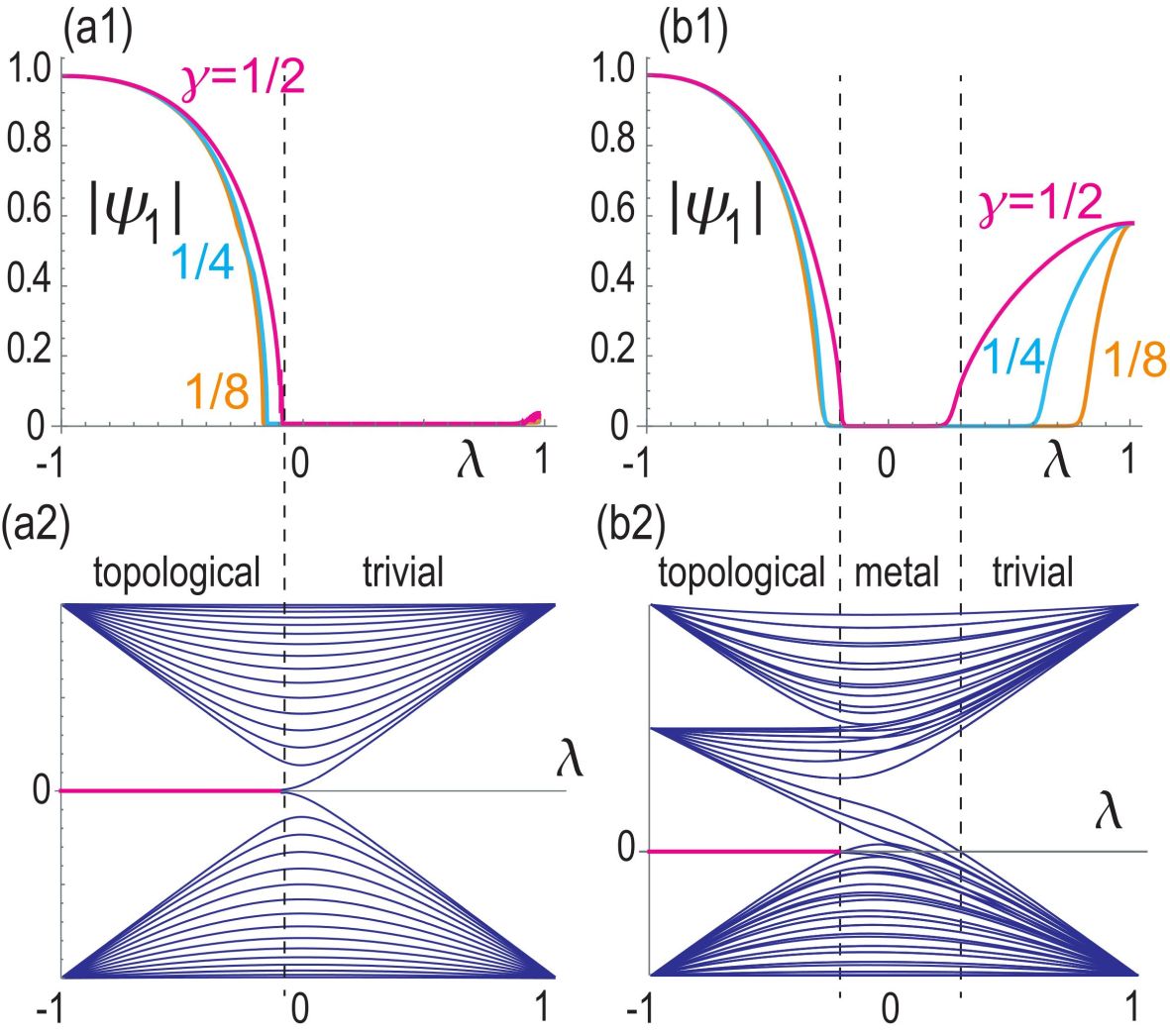}}
\caption{Saturated amplitude $|\protect\psi _{1}|$ at (a1) the edge site in
the SSH model, and (b1) the corner site in the breathing Kagome model, where 
$\protect\gamma =1/2$ for magenta, $\protect\gamma =1/4$ for cyan and $%
\protect\gamma =1/8 $ for orange colors. (a2) The energy spectrum in the SSH
model made of a finite chain with topological edge states in red. (b2) The
energy spectrum of the breathing Kagome model in triangle geometry with
topological corner states in red. The horizontal axis is $\protect\lambda $.
We have set $\kappa =1$, $\protect\eta =1$ and $\protect\xi =2$. }
\label{FigDistribute}
\end{figure}

\subsection{Quench dynamics}

Quench dynamics starting from one site is a good signal to detect whether
the system is topological or trivial\cite{QWalk}. It is also applicable to
various nonlinear systems\cite{TopoToda,MechaRot,NLPhoto,Sine}. Let us study
a quench dynamics starting from one site indexed by $m$,%
\begin{equation}
\psi _{n}\left( t\right) =\delta _{n,m}\quad \text{at}\quad t=0,
\label{IniCon}
\end{equation}%
where $m=1$ represents the left-edge site or the top-corner site as in Fig.%
\ref{FigKagomeIllust}(a2) and (b2).

\begin{figure}[t]
\centerline{\includegraphics[width=0.48\textwidth]{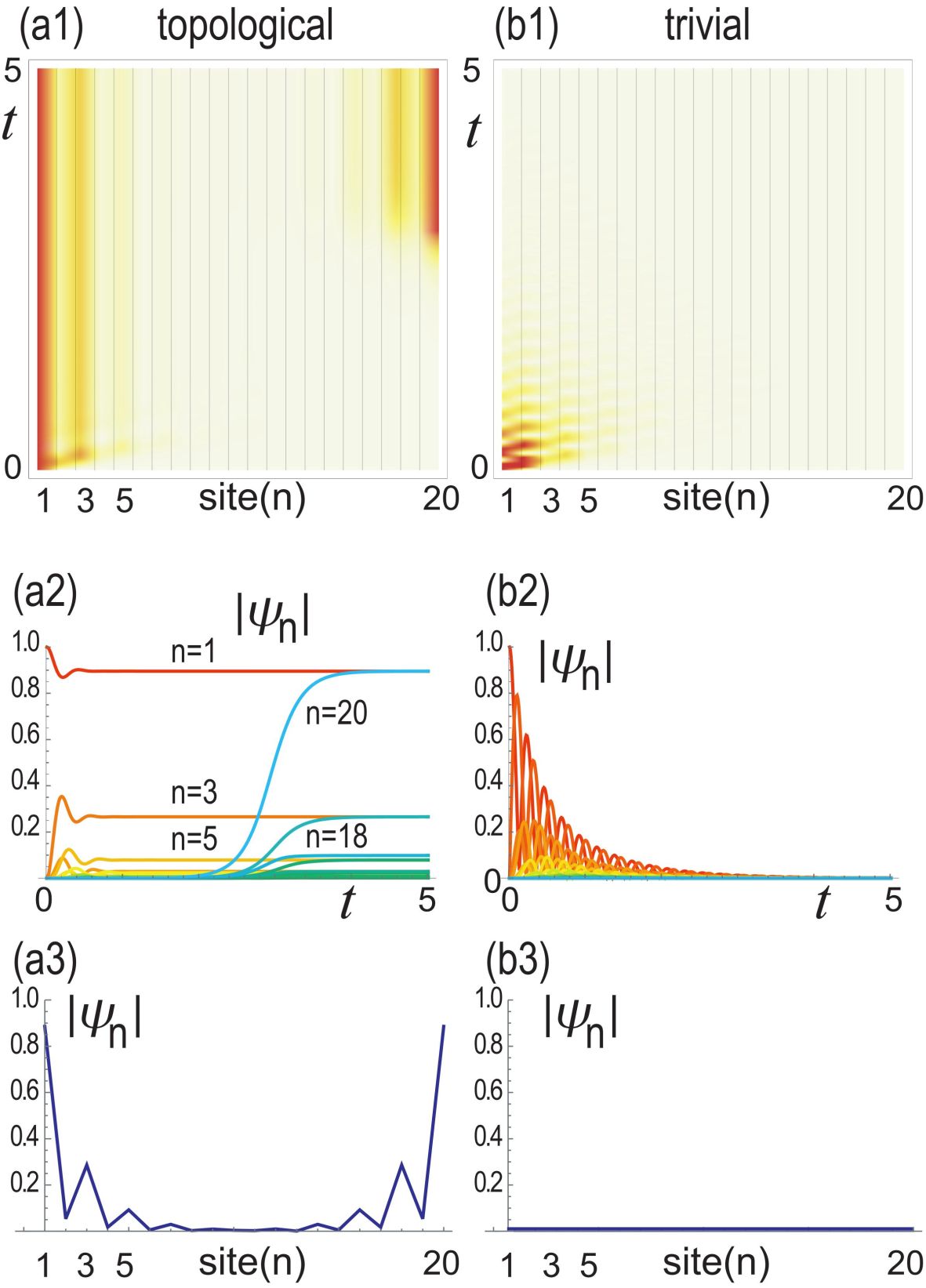}}
\caption{Density plot of the time evolution of the amplitude $|\protect\psi %
_{n}|$ as a function of $n$\ in (a1) the topological phase with $\lambda =-0.5$ and (b1) the
trivial phase with $\lambda =0.5$ in the SSH model. The amplitude $|\protect\psi _{n}|$ as a
function of $t$\ for various $n$ in (a2) the topological phase and (b2) the
trivial phase, where red curves indicate $n=1$, while cyan curves indicate $%
n=N$ in (a2) and (b2). The saturated amplitude $|\protect\psi _{n}|$ as a
function of $n$ in (a3) the topological phase and (b3) the trivial phase
after enough time. We have set $\kappa =1$, $\protect\eta =1$, $\protect\gamma =1/2$ and $%
\protect\xi =2$. We take a sample with $N=20$.}
\label{FigSSHDensity}
\end{figure}

\section{Nonlinear non-Hermitian SSH model}

\subsection{Model}

A topological laser based on the SSH model has been discussed\cite%
{Schome,Weimann,Jean,Ota18,Parto,Zhao,Malzard,MalzardOpt}. We consider the
case where the hopping matrix is governed by the SSH matrix in Eq.(\ref{DSG}%
). The matrix is explicitly given by%
\begin{eqnarray}
M_{nm} &=&-\left( \kappa _{A}+\kappa _{B}\right) \delta _{n,m}+\kappa
_{A}\left( \delta _{2n,2m-1}+\delta _{2m,2n-1}\right)  \notag \\
&&+\kappa _{B}\left( \delta _{2n,2m+1}+\delta _{2m,2n+1}\right) .
\label{HoppiMatrix}
\end{eqnarray}%
The explicit equations for a finite chain with length $N$\ are given by%
\begin{eqnarray}
i\frac{d\psi _{2n-1}}{dt} &=&\kappa _{A}\left( \psi _{2n}-\psi
_{2n-1}\right) +\kappa _{B}\left( \psi _{2n-2}-\psi _{2n-1}\right)  \notag \\
&&-i\gamma \left( 1-\xi \frac{\delta _{n,1}}{1+\left\vert \psi
_{2n-1}\right\vert ^{2}/\eta }\right) \psi _{2n-1},  \label{SSH1} \\
i\frac{d\psi _{2n}}{dt} &=&\kappa _{B}\left( \psi _{2n+1}-\psi _{2n}\right)
+\kappa _{A}\left( \psi _{2n-1}-\psi _{2n}\right)  \notag \\
&&-i\gamma \left( 1-\xi \frac{\delta _{n,N}}{1+\left\vert \psi
_{2n}\right\vert ^{2}/\eta }\right) \psi _{2n}.  \label{SSH2}
\end{eqnarray}%
It is convenient to introduce the coupling strength $\kappa $\ and the
dimerization parameter $\lambda $ by 
\begin{equation}
\kappa _{A}=\kappa \left( 1+\lambda \right) ,\quad \kappa _{B}=\kappa \left(
1-\lambda \right) ,  \label{DimerParam}
\end{equation}%
with $|\lambda |\leq 1$. The isolated edge limit is realized at $\lambda =-1$%
\ in Fig.\ref{FigKagomeIllust}(a1).

The present SSH model (\ref{DSG}) has the same topological structure as in
the original SSH model by the following reasoning. First, the contribution
of the nonlinear gain term ($\propto \gamma \xi $) is negligible in the bulk
since it exists only at the edge site. Next, the loss term ($\propto \gamma $%
) only shift the matrix as%
\begin{equation}
\overline{M}_{nm}\equiv M_{nm}-i\gamma \delta _{nm}.
\end{equation}%
Eq.(\ref{DSG}) is rewritten in the form of the linear model for the bulk ($%
n\neq \bar{n}$),%
\begin{equation}
i\frac{d\psi _{n}}{dt}=\sum_{nm}\overline{M}_{nm}\psi _{m}.
\label{HoppiMatrixA}
\end{equation}%
Then, the topological properties are determined by the matrix $\overline{M}%
_{nm}$, and they are identical to those in the original SSH model, as
explained in Appendix: See Eq.(\ref{ChiralIndexA}). We show the band
structure for a finite chain in Fig.\ref{FigDistribute}(a2). The system is
topological for $\lambda <0$ with the emergence of the topological edge
states \ marked in red, while it is trivial for $\lambda >0$.

\begin{figure}[t]
\centerline{\includegraphics[width=0.48\textwidth]{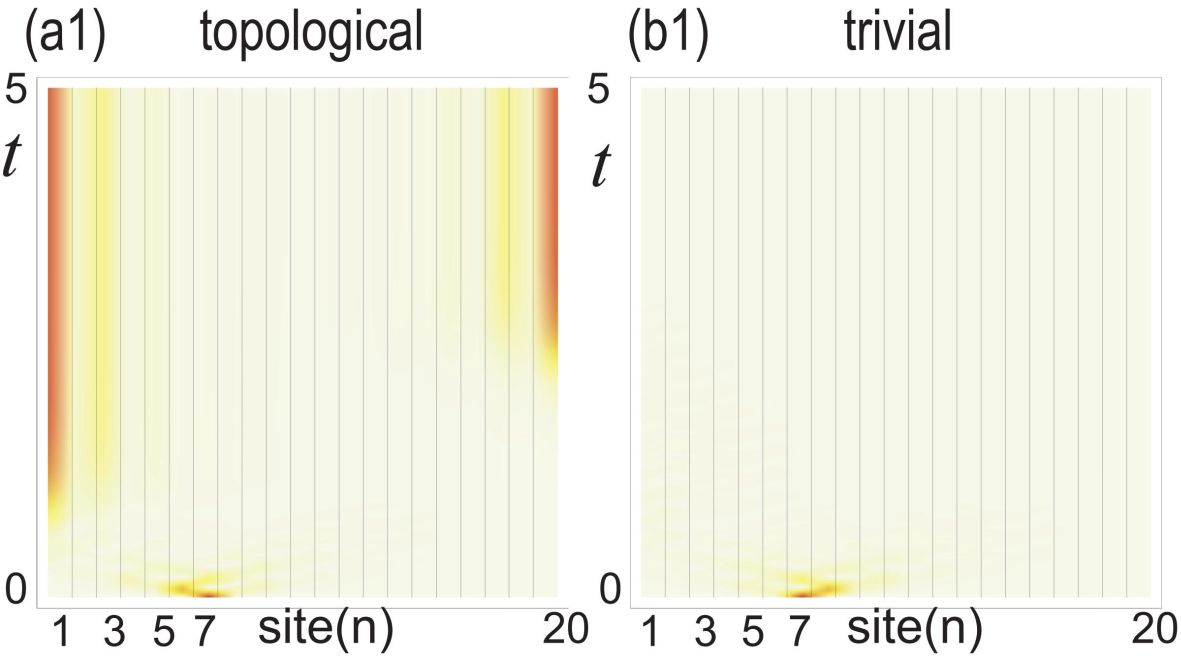}}
\caption{Density plot of the time evolution of the amplitude $|\protect\psi %
_{n}|$ as a function of $n$\ in (a1) the topological phase with $\lambda =-0.5$ and (b1) the
trivial phase  with $\lambda =0.5$ in the SSH model. We start from the site $n=7$. We have set $\kappa =1$, $\protect\eta =1$, $\protect\gamma =1/2$ and $%
\protect\xi =2$. We take a sample with $N=20$.}
\label{FigInter}
\end{figure}

\subsection{Quench dynamics}

We numerically solve Eqs.(\ref{SSH1}) and (\ref{SSH2}) under the initial
condition (\ref{IniCon}) by taking $m=1$. We show the time evolution of $%
\left\vert \psi _{n}\right\vert $ in Fig.\ref{FigSSHDensity}. The quench
dynamics is significantly different between the topological and trivial
phases. In the topological phase, the amplitude $|\psi _{1}|$ at the left
edge is rapidly saturated as shown in the red curve in Fig.\ref%
{FigSSHDensity}(a2), while there is a delay in the saturation of the
amplitude $|\psi _{N}|$ at the right edge as shown in the cyan curve in Fig.%
\ref{FigSSHDensity}(a2). The delay implies the propagation of a wave along a
chain, although the propagation of the wave is invisible in Fig.\ref%
{FigSSHDensity}(a1), as is consistent with Eq.(\ref{Decay}). It means that a
wave with a tiny amplitude transfers the information of the excitation to
the right edge and induces a stimulated emission. On the other hand, in the
trivial phase, the amplitude $|\psi _{1}|$ rapidly decreases as shown in Fig.%
\ref{FigSSHDensity}(a2). Furthermore, there is no excitation $|\psi _{N}|$
at the right edge.

Fig.\ref{FigSSHDensity}(a3) shows a spatial profile of the amplitude $%
\left\vert \psi _{n}\right\vert $\ after enough time, which is the DOS for a
pair of topological edge states in nonlinear non-Hermitian system.

We find that there is no reflection of the propagating wave by the right
edge. This is due to the loss term ($\propto \gamma $) in the bulk. It is
highly contrasted with the case of the Hermitian model. In addition, the
amplitudes $\left\vert \psi _{1}\right\vert $\ and $\left\vert \psi
_{N}\right\vert $\ are always identical. It is due to reflection symmetry $%
x\longleftrightarrow N-x$ in the right-hand side of Eqs.(\ref{SSH1}) and (%
\ref{SSH2}) because $d\psi _{n}/dt=0$\ for the stationary solution. This is
confirmed in Eq.(\ref{stationary}) explicitly for the limit ($\lambda =-1$)
of the isolated edge states, and\ numerically for any value of $\lambda $.

We show the saturated amplitude $|\psi _{1}|$ as a function of the
dimerization $\lambda $ in Fig.\ref{FigDistribute}(a1). It is finite for the
topological phase although it deviates from 1 other than $\lambda =-1$ due
to the hopping term. On the other hand, it is almost zero for the trivial
phase. These features correspond to the emergence or the absence of the
topological edge states as shown in Fig.\ref{FigDistribute}(a2). Namely, the
quench dynamics well signatures the topological phase transition although
there are nonlinear non-Hermitian terms.

We also study the dynamics under the initial condition (\ref{IniCon}) by
taking the site $m$ in the bulk. The result is shown in Fig.\ref{FigInter}
by choosing $m=7$. All sites belonging to the two topological edge states
are stimulated after a delay with the intensity depending on the DOS. The
timing of the stimulation is determined by the distance from the initial
site.

\begin{figure}[t]
\centerline{\includegraphics[width=0.48\textwidth]{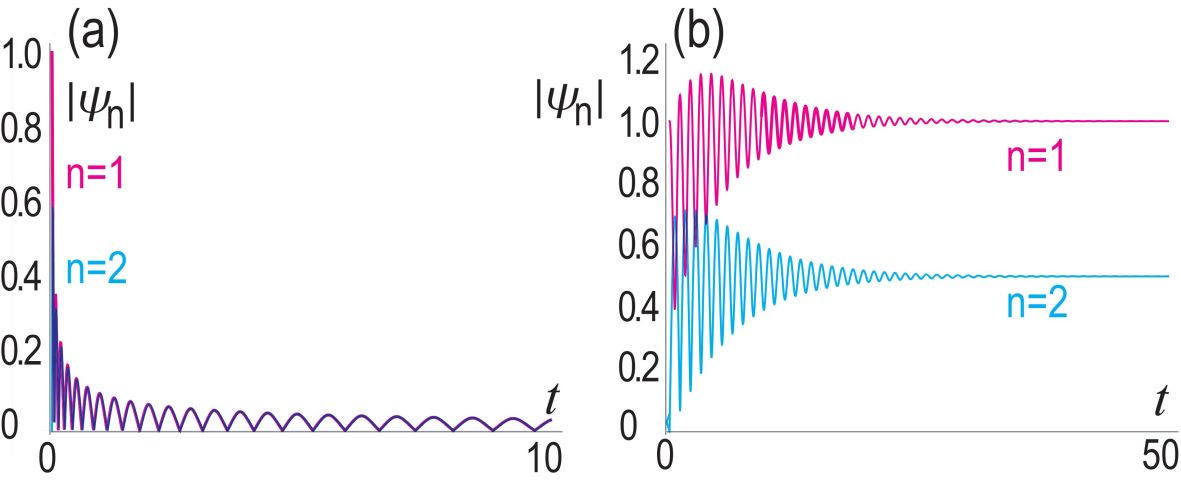}}
\caption{(a) Dynamics of the amplitude $|\protect\psi _{n}|$ of the dimer in
the SSH model (\protect\ref{dimer1}) and (\protect\ref{dimer2}). (b) The
dynamics of the trimer in the breathing Kagome model (\protect\ref{trimer1})
and (\protect\ref{trimer2}). We have set $\protect\kappa _{A}=1$, $\protect%
\eta =1$, $\protect\gamma =1/2$ and $\protect\xi =2$.}
\label{FigDimer}
\end{figure}

\begin{figure*}[t]
\centerline{\includegraphics[width=0.88\textwidth]{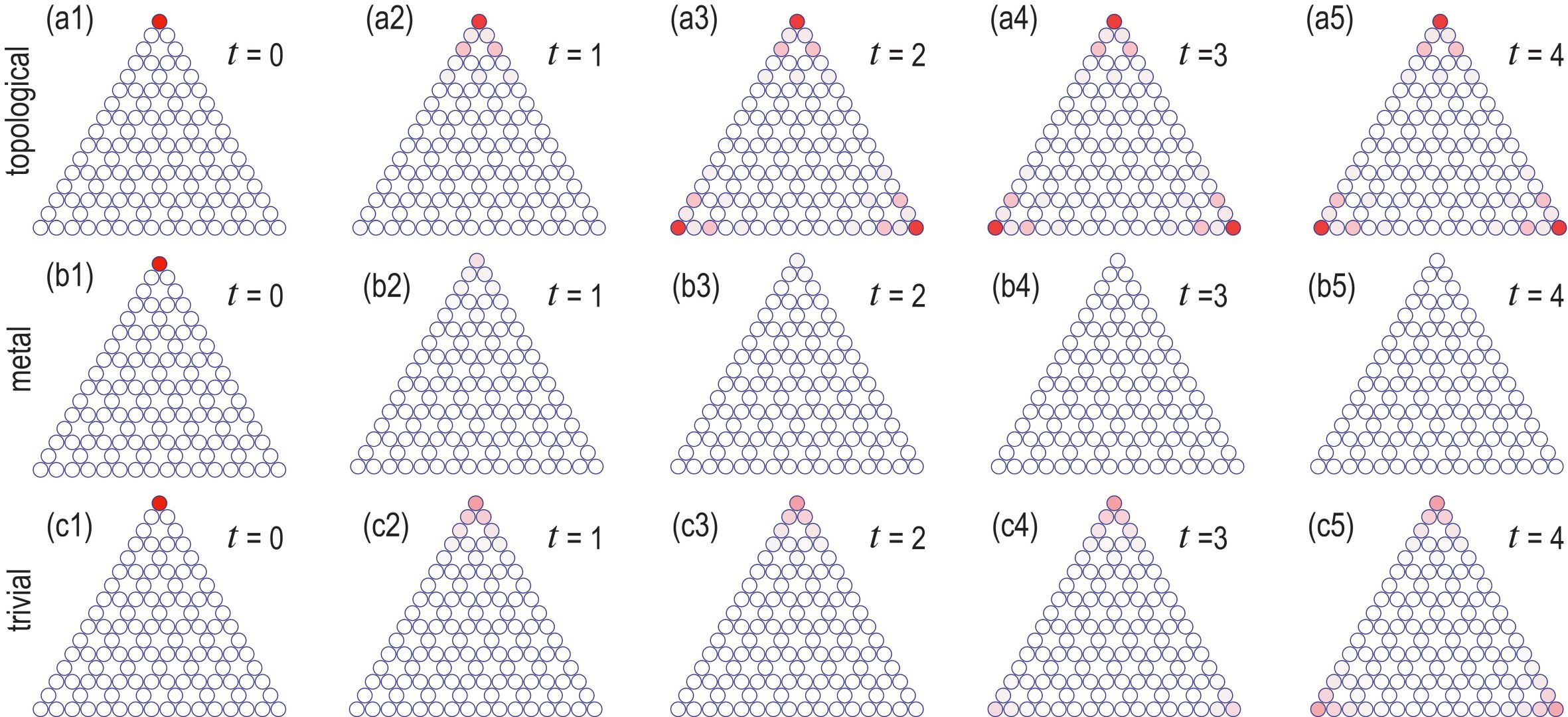}}
\caption{Time evolution of the amplitude $\left\vert \protect\psi %
_{n}\right\vert $ of the breathing Kagome model in (a) the topological phase
with $\protect\lambda =-0.5$, (b) the metal phase with $\protect\lambda =0.2$
and (c) the trivial phase with $\protect\lambda =0.5$. The color density
indicates the amplitude $|\protect\psi _{n}|$. (a5) and (c5) give the
profiles of the topological corner states and the trimer states together
with their DOS, respectively. We have set $\kappa =1$, $\protect\eta =1$, $\protect\gamma %
=1/2$ and $\protect\xi =2$. We take a triangle with $N=108$. }
\label{FigKagomeDynamics}
\end{figure*}

\subsection{Dimer states}

We study the trivial phase ($\lambda >1$). In particular, we may solve the
equations of motion analytically in the dimer limit ($\lambda =1$). In this
case, we obtain a closed set of equation for the sites $1$ and $2$,%
\begin{align}
i\frac{d\psi _{1}}{dt}& =\kappa _{A}\left( \psi _{2}-\psi _{1}\right)
-i\gamma \left( 1-\xi \frac{1}{1+\left\vert \psi _{1}\right\vert ^{2}/\eta }%
\right) \psi _{1},  \label{dimer1} \\
i\frac{d\psi _{2}}{dt}& =\kappa _{A}\left( \psi _{1}-\psi _{2}\right) .
\label{dimer2}
\end{align}%
The stationary solutions of (\ref{dimer1}) and (\ref{dimer2}) are either the
trivial one%
\begin{equation}
\psi _{1}=\psi _{2}=0,
\end{equation}%
or a nontrivial one%
\begin{equation}
\psi _{1}=\sqrt{\eta \left( \xi -1\right) },\qquad \psi _{2}=\text{const}.
\label{dimer3}
\end{equation}%
We note that the right-hand side of Eq.(\ref{dimer2}) is not necessary to be
zero because only the phase rotates if it is not zero. It is a dynamical
problem depending on the initial condition which stationary solution is
actually chosen.

We show numerical solution with the initial condition $\psi _{1}=1$ and $%
\psi _{2}=0$ in Fig.\ref{FigDimer}(a). The amplitudes $\left\vert \psi
_{1}\right\vert $ and $\left\vert \psi _{2}\right\vert $ exponentially
decreases to zero with oscillations. Hence, the stationary solutions are $%
\psi _{1}=\psi _{2}=0$. Furthermore, we have found numerically that $%
\left\vert \psi _{1}\right\vert $ is almost zero in the trivial phase as in
Fig.\ref{FigDistribute}(a). As far as we have checked numerically, there is
no nontrivial dimer solution in the trivial phase of the SSH model.

\section{Nonlinear non-Hermitian breathing Kagome model}

\subsection{Quench dynamics}

We proceed to investigate the system where the hopping matrix $M_{nm}$
describes the breathing Kagome lattice, whose lattice structure is
illustrated in Fig.\ref{FigKagomeIllust}(b). The hopping matrix and the
topological number are given by (\ref{H3}) and (\ref{PolarP}) in Appendix.
There are topological, trivial and metal phases in the breathing Kagome model%
\cite{EzawaKagome} as shown in Fig.\ref{FigDistribute}(b2).

By solving Eq.(\ref{DSG}) under the initial condition (\ref{IniCon}) with
the choice of $m=1$, we show the time evolution of the amplitude $\left\vert
\psi _{n}\right\vert $ in Fig.\ref{FigKagomeDynamics} and in Fig.\ref%
{FigKagomeDensity}, where $n=1$ denotes the top-corner site. Fig.\ref%
{FigKagomeDynamics} displays a global picture how the stimulated signal at
site $n=1$ propagates all over the sites as time evolves. Fig.\ref%
{FigKagomeDensity} displays a detailed evolution along the left side of the
lattice.\ We also show the saturated amplitude as a function of $\lambda $\
in Fig.\ref{FigDistribute}(b1). The saturated amplitudes are identical at
the three corner sites due to trigonal symmetry of the breathing Kagome
lattice, as found in Fig.\ref{FigKagomeDynamics}.

We comment that Fig.\ref{FigKagomeDynamics}(a5) shows a spatial profile of
the saturated amplitude $\left\vert \psi _{n}\right\vert $, which is the DOS
for three topological corner states in nonlinear non-Hermitian system.

\begin{figure}[t]
\centerline{\includegraphics[width=0.48\textwidth]{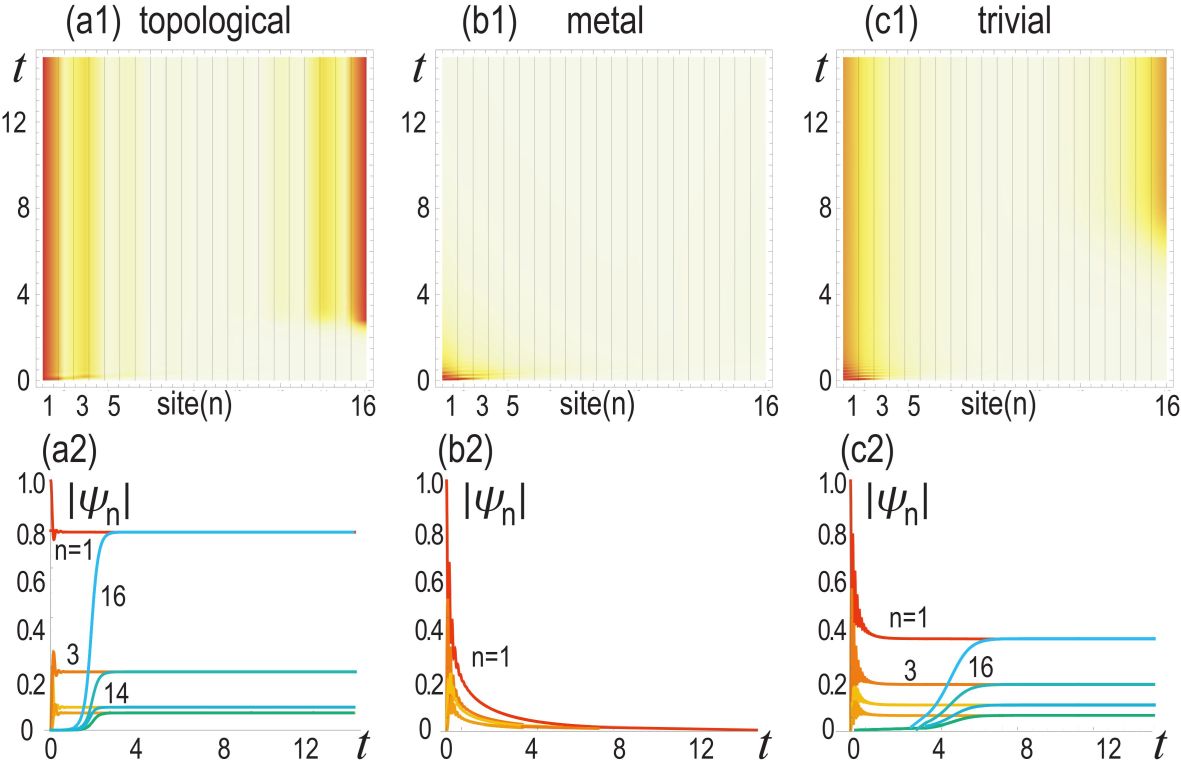}}
\caption{Density plot of the time evolution of the amplitude $|\protect\psi %
_{n}|$ along one side of the breathing Kagome lattice in (a1) topological
phase with $\protect\lambda =-0.5$, (b1) metal phase with $\protect\lambda %
=0.2$, and (c1) trivial phase with $\protect\lambda =0.5$. Time evolution of
the amplitude $\left\vert \protect\psi _{n}\right\vert $ of the breathing
Kagome model for various $n$ in (a2) topological phase with $\protect\lambda %
=-0.5$, (b2) metal phase with $\protect\lambda =0.2$, and (c2) trivial phase
with $\protect\lambda =0.5$. We have set $\kappa =1$, $\protect\eta =1$, $\protect\gamma %
=1/2$ and $\protect\xi =2$. We take a triangle with $N=108$.}
\label{FigKagomeDensity}
\end{figure}

These figures show typically different behaviors in the topological, trivial
and metal phases. First, in the topological phase, the amplitudes at the
other two corner sites increase after a delay to the saturated amplitude as
shown in Fig.\ref{FigKagomeDensity}(a2). Second, in the metal phase, the
amplitude at the corner site rapidly decreases as in Fig.\ref%
{FigKagomeDensity}(b2). These features are very much similar to those in the
the SSH model.

On the other hand, in the trivial phase, there is a significant difference
between the SSH model and breathing Kagome model. In the SSH model, the
amplitude rapidly decreases as in Fig.\ref{FigSSHDensity}(b2). However, this
is not the case for the breathing Kagome model. The amplitude at the top
corner site suddenly decreases to the saturated value, while those at the
bottom corner sites increase to the saturated value after a delay as in Fig.%
\ref{FigKagomeDensity}(c2). Furthermore, the saturated amplitude $\left\vert
\psi _{1}\right\vert $ depends on the magnitude of $\xi $ as shown in Fig.%
\ref{FigDistribute}(b1). We pursue the reason why the amplitudes do not
vanish in the trivial phase. We will see that it is due to the formation of
the trimer state.

\subsection{Trimer states}

We consider the limit $\lambda =1$, where the system is decomposed into a
set of trimers as in Fig.\ref{FigKagomeIllust}(b3). The equations of motion
is explicitly given by%
\begin{align}
i\frac{d\psi _{1}}{dt}& =\kappa _{A}(\psi _{2}+\psi _{3})-i\gamma \left(
1-\xi \frac{1}{1+\left\vert \psi _{1}\right\vert ^{2}/\eta }\right) \psi
_{1}, \\
i\frac{d\psi _{2}}{dt}& =\kappa _{A}\left( \psi _{1}+\psi _{3}\right) , \\
i\frac{d\psi _{3}}{dt}& =\kappa _{A}\left( \psi _{1}+\psi _{2}\right) .
\end{align}%
Without loss of generality we may set $\psi _{2}=\psi _{3}$ and obtain%
\begin{align}
i\frac{d\psi _{1}}{dt}& =2\kappa _{A}\psi _{2}-i\gamma \left( 1-\xi \frac{1}{%
1+\left\vert \psi _{1}\right\vert ^{2}/\eta }\right) \psi _{1},
\label{trimer1} \\
i\frac{d\psi _{2}}{dt}& =\kappa _{A}\left( \psi _{1}+\psi _{2}\right) .
\label{trimer2}
\end{align}%
The stationary solutions are either the trivial one%
\begin{equation}
\psi _{1}=\psi _{2}=0,
\end{equation}%
or a nontrivial one%
\begin{equation}
\psi _{1}=\sqrt{\eta \left( \xi -1\right) },\qquad \psi _{2}=\text{const},
\label{trimerA}
\end{equation}%
which is identical to the stationary solution (\ref{dimer3}) in the SSH
model. There is no $\gamma $ dependence in the stationary solution (\ref%
{trimerA}), which agrees with in Fig.\ref{FigDistribute}(b1).

We have found analytically the trimer state in the limit of $\lambda =1$ in
the trivial phase. A numerical solution is given as a function of time $t$
in Fig.\ref{FigDimer}(b), where $\left\vert \psi _{1}\right\vert $ and $%
\left\vert \psi _{2}\right\vert $ approach two saturated values (\ref%
{trimerA}), as is consistent with Fig.\ref{FigDistribute}(b1). Fig.\ref%
{FigDistribute}(b1) suggests that the trimer state is formed also
away from the limit $\lambda =1$\ depending on the value of $\gamma $.
Indeed, Fig.\ref{FigKagomeDynamics}(c5) shows a spatial profile of 
three trimer states at $\lambda =0.5$. The formation of trimer states
implies the presence of a nontopological laser in the breathing Kagome
model. These behaviors are contrasted with the dimer state in the SSH model,
where $\left\vert \psi _{1}\right\vert =0$ in the trivial phase as in Fig.%
\ref{FigDistribute}(a1).

\section{Discussion}

A topological laser provides a unique arena of topology, non-Hermicity and
nonlinearity. We have studied topological lasing in the SSH lattice and the
breathing Kagome lattice. They have topological edge or corner states in the
topological phase. When any one site is stimulated, all sites belonging to
the topological edge or corner states begin to emit stable laser light
depending on the DOS. The results would be universal for the physics of
topological lasing with the use of topological edge or corner states.

Non-Hermicity and nonlinearity play essential roles for the stabilization of
lasing. Without the nonlinearity, there is no stable lasing because the
amplitude exponentially grows or decays. The amplifier plays an essential
role of the lasing. If there is no loss term, there should be reflection
wave at the right edge site or the bottom corner sites, which are absent in
a topological laser.

On the other hand, nontopological lasing is not universal. Indeed, the
quench dynamics is different between the SSH model and the breathing Kagome
model in the trivial phase. There is no stable laser emission in the SSH
model. However, once one site is stimulated, all three corner states emit
stable laser lights in the breathing Kagome model due to the formation of
trimer states at the three corners.

In conclusion, we have found stable laser emission to occur in the
topological phase. Our results show that topological lasers provide an ideal
play ground of nonlinear non-Hermitian topological physics, where the
profile of a topological edge or corner state is observable by measuring the
intensity of lasing.

The author is very much grateful to N. Nagaosa for helpful discussions on
the subject. This work is supported by the Grants-in-Aid for Scientific
Research from MEXT KAKENHI (Grants No. JP17K05490 and No. JP18H03676). This
work is also supported by CREST, JST (JPMJCR16F1 and JPMJCR20T2).

\appendix

\section{Topological number in the SSH model}

The hopping matrix of the SSH model (\ref{HoppiMatrix}) is given by%
\begin{equation}
\overline{M}\left( k\right) =-\left( \kappa _{A}+\kappa _{B}+i\gamma \right)
I_{2}+M_{0}\left( k\right)  \label{MkBar}
\end{equation}%
in the momentum space, with%
\begin{equation}
q\left( k\right) =\kappa _{A}+\kappa _{B}e^{-ik},
\end{equation}%
and%
\begin{equation}
M_{0}\left( k\right) =\left( 
\begin{array}{cc}
0 & q\left( k\right) \\ 
q^{\ast }\left( k\right) & 0%
\end{array}%
\right) .  \label{Mk0}
\end{equation}%
The topological number in the original SSH model is given by the Berry phase%
\begin{equation}
\Gamma =\frac{1}{2\pi }\int_{0}^{2\pi }A\left( k\right) dk,
\label{ChiralIndexA}
\end{equation}%
where $A\left( k\right) =-i\left\langle \phi (k)\right\vert \partial
_{k}\left\vert \phi (k)\right\rangle $ is the Berry connection with $\phi
(k) $ the eigenfunction of $\overline{M}\left( k\right) $. Note that the
diagonal term in Eq.(\ref{HoppiMatrixA}) with Eq.(\ref{HoppiMatrix}) does
not contribute to the topological charge because the wave function $\phi (k)$
does not depend on the diagonal term. Hence, the present model (\ref{DSG})
has the same phases as the original SSH model. The system is topological for 
$\lambda <0$ and trivial for $\lambda >0$.

\section{Breathing Kagome model}

The hopping matrix of the breathing Kagome model is given by\cite%
{EzawaKagome} 
\begin{equation}
M\left( \mathbf{k}\right) =-\left( 
\begin{array}{ccc}
0 & h_{12} & h_{13} \\ 
h_{12}^{\ast } & 0 & h_{23} \\ 
h_{13}^{\ast } & h_{23}^{\ast } & 0%
\end{array}%
\right)  \label{H3}
\end{equation}%
with 
\begin{align}
h_{12}& =\kappa _{A}e^{i\left( k_{x}/2+\sqrt{3}k_{y}/2\right) }+\kappa
_{B}e^{-i\left( k_{x}/2+\sqrt{3}k_{y}/2\right) }, \\
h_{23}& =\kappa _{A}e^{i\left( k_{x}/2-\sqrt{3}k_{y}/2\right) }+\kappa
_{B}e^{i\left( -k_{x}/2+\sqrt{3}k_{y}/2\right) }, \\
h_{13}& =\kappa _{A}e^{ik_{x}}+\kappa _{B}e^{-ik_{x}}
\end{align}%
in the momentum space, where we have introduced two hopping parameters $%
\kappa _{A}$ and $\kappa _{B}$ corresponding to upward and downward
triangles in Fig.\ref{FigKagomeIllust}(b).

\section{Topological number in the breathing Kagome model}

The topological number in the breathing Kagome model is defined by\cite%
{EzawaKagome} 
\begin{equation}
\Gamma =3\left( p_{x}^{2}+p_{y}^{2}\right) ,
\end{equation}%
where 
\begin{equation}
p_{i}=\frac{1}{S}\int_{\text{BZ}}A_{i}d^{2}\mathbf{k},  \label{PolarP}
\end{equation}%
with $A_{i}=-i\left\langle \phi (\mathbf{k})\right\vert \partial
_{k_{i}}\left\vert \phi (\mathbf{k})\right\rangle $ being the Berry
connection with $x_{i}=x,y$, and $S=8\pi ^{2}/\sqrt{3}$ being the area of
the Brillouin zone; $\phi (\mathbf{k})$ the eigenfunction of $M\left( 
\mathbf{k}\right) $. We obtain $\Gamma =0$ for $1\geq \lambda >1/3$, which
is the trivial phase with no topological corner states. On the other hand,
we obtain $\Gamma =1$ for $-1\leq \lambda <-1/3$, which is the topological
phase with the emergence of three topological corner states. Finally, $%
\Gamma $ is not quantized for $-1/3<\lambda <1/3$, which is the metal phase.

\end{document}